\documentstyle{mn}

\title[Horsehead Nebula]{SCUBA observations of the Horsehead Nebula
-- what did the horse swallow?}

\author[D. Ward-Thompson et al.]
{D. Ward-Thompson$^{1,2}$, D. Nutter$^1$, S. Bontemps$^2$,
A. Whitworth$^1$, R. Attwood$^1$ \\
$^1$Department of Physics and Astronomy, Cardiff University,
5 The Parade, Cardiff, CF24 3YB \\
$^2$Observatoire de Bordeaux, 2 rue de l'Observatoire, 33270 Floirac, France}

\date{Accepted 2006 March ?; received 2006 March 17; in original form
2005 December 6.}

\pagerange{000--000}

\begin{document}

\label{firstpage}

\maketitle
\begin{abstract}
We present observations taken with the Submillimetre Common-User Bolometer
Array (SCUBA) on the James Clerk Maxwell Telescope (JCMT) of the
Horsehead Nebula in Orion (B33), at wavelengths of 450 and 850~$\mu$m. 
We see bright emission from that part of the cloud associated
with the photon-dominated region (PDR)
at the `top' of the horse's head, which we label B33-SMM1.
We characterise the physical parameters of the extended
dust responsible for this emission, and
find that B33-SMM1 contains a more dense core than was previously 
suspected, with a mass of $\sim$2~M$_\odot$ in a region of 0.31 $\times$ 
0.13~pc, and a peak volume density of $\sim$6 $\times$ 10$^5$ cm$^{-3}$. 
We compare the SCUBA data with data from the Infrared Space Observatory (ISO)
and find that the emission at 6.75-$\mu$m is offset towards the west,
indicating that the mid-infrared emission is tracing the PDR while
the submillimetre emission comes from the molecular cloud core behind 
the PDR. We calculate the virial balance of this core and find
that it is not gravitationally bound but is being confined by the
external pressure from the HII region IC434, and that it will either
be destroyed by the ionising radiation, or else may
undergo triggered star formation.

Furthermore we find evidence for a lozenge-shaped clump in the `throat' of 
the horse, which is not seen in emission at shorter wavelengths. We 
label this source B33-SMM2 and find that it is brighter at submillimetre 
wavelengths than B33-SMM1. We calculate the physical parameters of SMM2 and
find it has a mass of $\sim$4~M$_\odot$ in a region 0.15 $\times$ 0.07~pc,
with a peak volume density of $\sim$2 $\times$ 10$^6$ cm$^{-3}$
and peak column density of $\sim$9 $\times$ 10$^{22}$ cm$^{-2}$.
SMM2 is seen in absorption in the 6.75-$\mu$m ISO data, from
which we obtain an independent estimate of the column density in excellent
agreement with that calculated from the submillimetre emission.
We calculate the stability of this core against collapse and 
find that it is in approximate gravitational virial equilibrium. This is
consistent with it being a pre-existing core in B33, possibly pre-stellar 
in nature, but that it may also eventually
undergo collapse under the effects of the HII region.
\end{abstract}

\section{Introduction}

The Horsehead Nebula in Orion is one of the most familiar images
in astronomy -- see Figure 1. It appears as B33 in the catalogue of dark
clouds of Barnard (1919). It has been observed by many people at many
wavelengths. Recent studies by Pound, Reipurth \& Bally (2003),
Abergel et al. (2003), Teyssier et al. (2004), Pety et al. (2005), and Habart 
et al. (2005) have concentrated on the edge of the nebula that is nearest 
to the HII region (the `top' of the horse's head). All derived various
values for the gas densities and temperatures across the photon-dominated
region (PDR) that exists along this edge of the nebula. 

Abergel et al. (2003) confirmed that the star $\sigma$ Orionis is responsible 
for the ionising radiation producing the PDR. They deduced that the PDR
is $\sim$0.01~pc in thickness and they found densities of $\sim$10$^4$
cm$^{-3}$ immediately behind the ionisation front. Teyssier et al. (2004)
presented observations of various carbon-bearing species, and Pety et al.
(2005) showed interferometer observations of the PDR from which they
concluded that no current model of PDRs could explain all of their data,
and that this might be explained by fragmentation of polycyclic aromatic
hydrocarbon (PAH) molecules by the intense UV field. Habart et al. (2005)
found very bright, narrow filaments of fluorescent H$_2$ emission along the 
edge of the PDR. Hily-Blant et al. (2005) studied the velocities across the 
nebula as a whole and found a complex series of helical rotational velocity 
gradients.

Pound et al. (2003) asserted that the Horsehead Nebula is an extension of the
L1630 dark cloud, seen against the background HII region IC434, illuminated
edge-on by the O9.5V star $\sigma$~Orionis. They
compared the nebula to the now equally famous Eagle nebula
with its `pillars' of gas and dust. They deduced that the two nebulae were 
formed in a somewhat similar fashion, by the action of nearby massive stars 
and their associated HII regions, either by an instability or an ionisation 
front. Williams, Ward-Thompson \& Whitworth (2001) argued that the formation 
of such `pillars' or `columns' of dust and gas is probably a fairly common 
phenomenon in star formation. 

\begin{figure}
\setlength{\unitlength}{1mm}
\noindent
\begin{picture}(80,85)
\put(0,0){\includegraphics{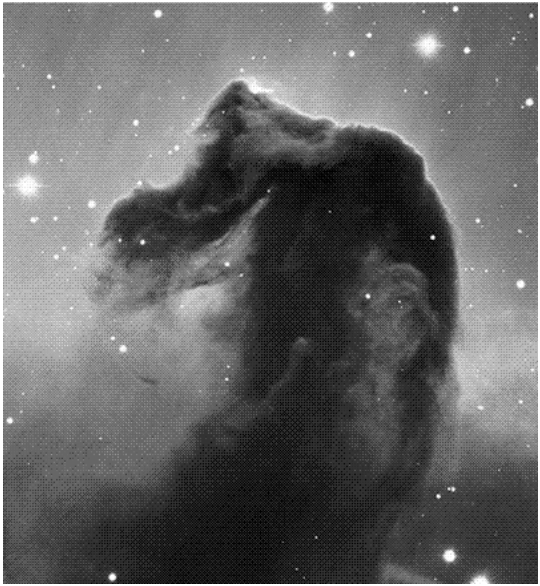}}
\end{picture}
\caption{Optical image of the Horsehead Nebula taken with the Very Large
Telescope (VLT). West is at the top, North
is to the left. Image courtesy of the European Southern Observatory (ESO).}
\end{figure}

One difference between the Eagle Nebula and
the Horsehead Nebula is that whereas dense condensations have been known
for some time to exist in the pillars of the Eagle Nebula (e.g. White et al. 
1999), until recently no similar condensations were known in the Horsehead 
Nebula. These dense condensations are the likely progenitors of the next 
generation of star formation and may give clues as to the initial conditions 
for star formation in such regions. Studies of these dense clumps may
therefore provide indicators that will allow us to differentiate between 
the various models of `triggered' and `spontaneous' star formation (e.g.
Ward-Thompson et al. 2006).

In this paper we present submillimetre continuum data of the Horsehead 
Nebula, retrieved from the James Clerk Maxwell Telescope (JCMT) data archive,
taken by the Submillimetre Common User Bolometer Array (SCUBA). Our goal
in obtaining the data was to search for condensations in the densest parts
of the Horsehead Nebula, to see whether future star formation is likely in
this region, before it is possibly destroyed in $\sim$5 $\times$ 10$^6$ years,
as estimated by Pound et al. (2003).

\begin{figure}
\setlength{\unitlength}{1mm}
\noindent
\begin{picture}(80,95)
\put(0,0){\includegraphics{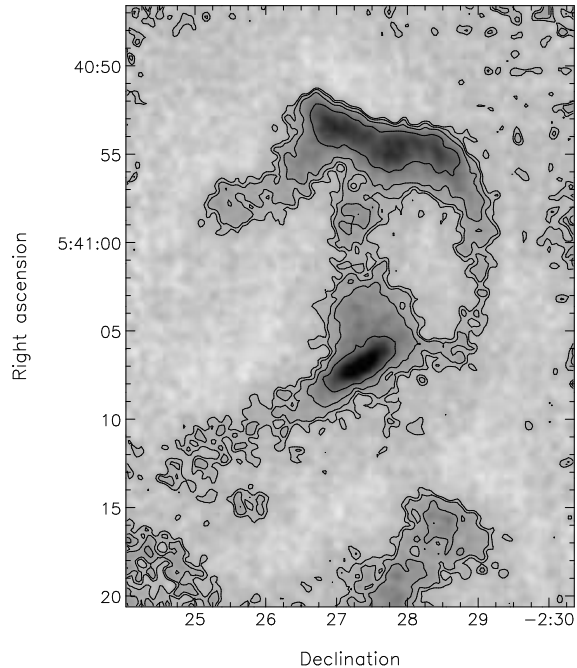}}
\end{picture}
\caption{SCUBA 850-$\mu$m image of the Horsehead Nebula. 
The region shown is approximately the same as that in Figure 1. 
West is at the top, North is to the left. Contour levels are at
2, 3, 5 \& 10~$\sigma$, where 1~$\sigma$ is 17~mJy/beam. 
The FWHM beamsize is 15 arcsec. The familiar
horsehead shape can be seen in this image, as well as the source SMM1
associated with
the PDR along the `top' of the horse's head, and the source SMM2
in the horse's `throat'.}
\end{figure}

\section{Data}

The data were obtained from the JCMT data archive. This source has been
observed several times by SCUBA. Only those data with
the highest signal-to-noise ratio were used to make
maps of the region at 850 and 450~$\mu$m. These were taken on 1999
November 30 at 02:49--04:35 hours HST (UT 12:49--14:35).
Data were taken at both wavelengths simultaneously using SCUBA in its
standard on-the-fly mapping mode (Holland et al. 1999).
Average sky conditions during the observations were
determined using the `skydip' method
and by comparison with the 1.3-mm sky opacity.
The 1.3mm optical depth ranged from 0.033 to 0.035 during the observations.
The 850-$\mu$m zenith optical depth was 0.117--0.123
corresponding to a mean zenith transmission of $\sim$90\%.
The 450-$\mu$m optical depth was 0.50--0.54
corresponding to a zenith transmission of $\sim$60\%.

The data were reduced in the normal way using the SCUBA User
Reduction Facility (Jenness \& Lightfoot 2000).
Calibration was performed using observations of the planet Mars.
We estimate that the absolute calibration uncertainty
is $\pm$10\% at 850 $\mu$m and $\pm$25\% at 450$\mu$m, based on
the consistency and reproducibility of the calibration.
The average beam size full-width at half maximum (FWHM)
was found to be 15 arcsec at 850 $\mu$m and 10 arcsec at 450 $\mu$m.
These numbers are both slightly higher than the notional 
beamwidths at these wavelengths. Fitting two-component gaussians to
the calibrator source showed there to be a negligible error beam
at 850~$\mu$m, 
but at 450 $\mu$m a significant error beam was detected that
was found to contribute up to 10\% of the flux density. 
This was taken into account in calibrating the data by using identical
apertures when determining the calibration on the planet Mars
and measuring the flux densities of the sources.
In the final maps it was found that
the average 1$\sigma$ noise (off-source)
was 17 mJy/beam at 850 $\mu$m and 110 mJy/beam at 450 $\mu$m.

Figure 2 shows the 850-$\mu$m data of the region, rotated to the same 
orientation as the optical image in Figure 1. The outline of the horse's
head can be clearly seen in this image. In addition, a brighter region
along the `top' of the horse's head can be seen,
roughly coincident with the PDR
discussed in section 1 above, which we here label B33-SMM1.
However, we also see a second bright source
in the image, in the area of the horse's `throat', looking just as if the
horse has swallowed a large lump. We here label this source B33-SMM2.

Figure 3 shows the 450-$\mu$m image of the same region. The signal-to-noise
ratio of this image is not as high as that of the 850-$\mu$m image.
Nonetheless the approximate outline of
the nebula can still be seen. Once again the source near
the PDR is visible -- SMM1.
The source in the horse's throat can also be seen in this image -- SMM2.

\begin{figure}
\setlength{\unitlength}{1mm}
\noindent
\begin{picture}(80,95)
\put(0,0){\includegraphics{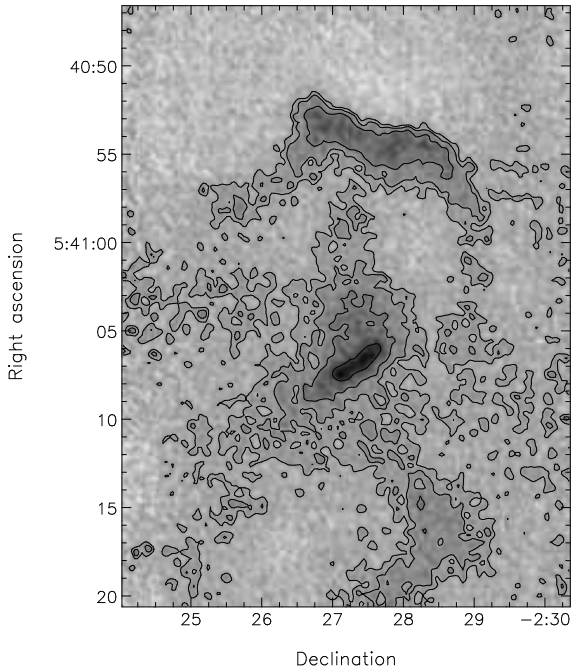}}
\end{picture}
\caption{SCUBA 450-$\mu$m image of the Horsehead Nebula. The same region
is shown as in Figure 2 and in the same orientation. Contour levels are
again at 2, 3, 5 \& 10~$\sigma$, where 1~$\sigma$ is now 110~mJy/beam.
The FWHM beamsize is 10 arcsec.
The signal-to-noise ratio is lower than
in the 850-$\mu$m image, although the horse's head
shape can still be made out. Once again the source associated with the
PDR, SMM1, and the source in the throat, SMM2, can be seen.}
\end{figure}

\begin{figure}
\setlength{\unitlength}{1mm}
\noindent
\begin{picture}(80,95)
\put(0,0){\includegraphics{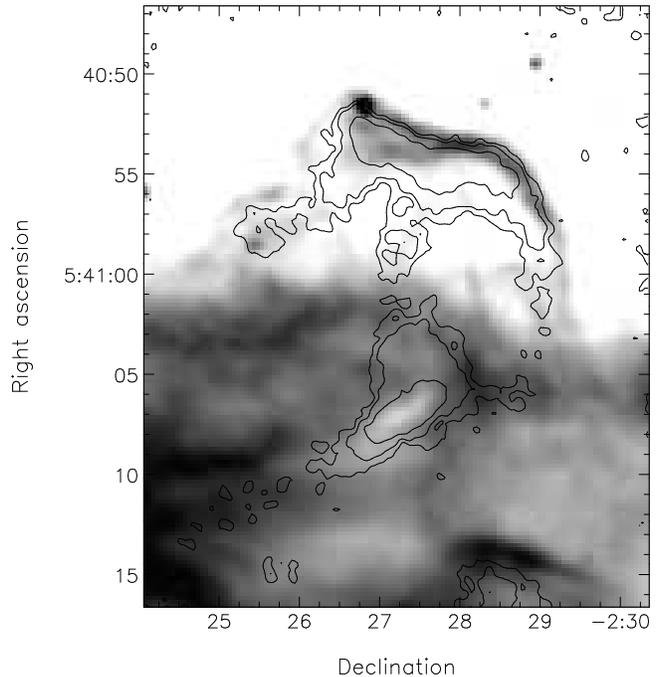}}
\end{picture}
\caption{SCUBA 850-$\mu$m isophotal contour map of the Horsehead Nebula
overlaid on a greyscale of the ISOCAM 6.75-$\mu$m image.
The region shown is the same as that in Figure~2. 
West is at the top, North is to the left. Contour levels are as
in Figure~2. The submillimetre emission from the source SMM1 can now be
seen to be offset from the infrared emission, with the latter lying
closer to the HII region and appearing almost to wrap around the
submillimetre source. The source SMM2 in the horse's throat can be
seen to be associated with a dip in the mid-infrared emission
(c.f. Abergel et al. 2003).}
\end{figure}

Table 1 lists the measured parameters of SMM1 \& 2. We quote the central 
positions and sizes of the two sources. The size of each source was
calculated by fitting the elliptical aperture that most closely matched 
the 3-sigma contour enclosing each source. The flux densities in these 
apertures were measured at 850 and 450~$\mu$m and are listed in Table 1, 
along with the peak flux density of each source.
We see from Table 1 that SMM1 is the more extended of the two sources,
although SMM2 is the brighter of the two, as
its peak flux density is greater at both wavelengths. Furthermore,
even though SMM2 is the more compact source, its
extended flux density is greater at 450~$\mu$m than SMM1, and it is
comparable at 850~$\mu$m. 

\begin{table*}
\caption{Positions and sizes, together with peak and integrated
flux densities (quoted to 3 significant figures), of the two submillimetre
sources B33-SMM1 \& 2. The peak flux densities are quoted in a single
beam, which was measured to be 15 arcsec at 850~$\mu$m and 10 arcsec at
450~$\mu$m. The integrated flux densities were measured within
the apertures quoted.
The 1-$\sigma$ error-bars on the peak flux densities
are 17 and 110 mJy/beam at 850 and 450~$\mu$m respectively. The absolute
error-bars are estimated to be $\pm$10\% and $\pm$25\% at 850 and 
450~$\mu$m respectively. The aperture sizes were defined 
by the best-fitting ellipse to the
3-$\sigma$ contour level enclosing each source.
The major and minor axes of the ellipse are quoted, together with
the position angle of the major axis, measured North through East.}
\begin{tabular}{lcccccccc}
\hline
Source & RA & Dec. & S$_{850}^{peak}$ & S$_{850}^{int}$ &
S$_{450}^{peak}$ & S$_{450}^{int}$ & Aperture (arcsec) &
Posn angle \\
 & (2000) & (2000) & (mJy/beam) & (mJy) & (mJy/beam) & (mJy) & (Major
$\times$ Minor) & (N $\rightarrow$ E) \\
\hline
B33-SMM1 & 05$^{\rm h}$ 40$^{\rm m}$ 54$^{\rm s}$
& $-$02$^\circ$ 27$^\prime$ 32$^{\prime\prime}$
& 280 & 1470 & 1050 & 6570 & 161 $\times$ 68 & 160$^\circ$ \\
B33-SMM2 & 05$^{\rm m}$ 41$^{\rm m}$ 07$^{\rm s}$
& $-$02$^\circ$ 27$^\prime$ 21$^{\prime\prime}$ 
& 430 & 1380 & 1550 & 10300 & 78 $\times$ 36 & 38$^\circ$ \\
\hline
\end{tabular}
\end{table*}

SMM1 appears to run roughly along the full length of the ridge of the PDR 
that has been extensively studied recently (e.g. Habart et al. 2005 and 
references therein). To check how well the extent of the submillimetre 
emission along the ridge matches the mid-infrared emission from this source,
as seen by the Infrared Space Observatory (ISO) using the ISOCAM mid-infrared 
camera, we compared the two datasets in more detail. 

Figure~4 shows once again the contours of 850-$\mu$m emission from 
Figure~2, but this time these are plotted over a grey-scale of the 6.75-$\mu$m 
emission seen by ISOCAM (c.f. Abergel et al. 2003). We see from this plot
that the mid-infrared emission lies along the western edge of
the 850-$\mu$m emission from SMM1. The infrared ridge is offset from
the submillimetre ridge by $\sim$20--25 arcsec. This is as expected,
given that the cloud is being externally heated by $\sigma$~Ori (e.g.
Pound et al. 2003). 

The warmer dust on the side of the cloud nearest to 
the heating source emits more strongly in the mid-infrared, while the
cooler dust further into the cloud emits at longer wavelengths. In fact
the mid-infrared emission appears to wrap right round the outside of
the cloud, exactly as expected for such an outside-in temperature gradient
(c.f. Ward-Thompson et al. 2002). In fact, the 850-$\mu$m emission matches
better the molecular line emission traced by various isotopomers of CO 
(c.f. Abergel et al. 2003; Habart et al. 2005).

SMM1 was seen as two separate sources by Johnstone, Matthews \& Mitchell
(2005), who used an automated source-finding procedure to locate all of 
the sources detected by SCUBA in the Orion region. We see SMM1 as
simply an undulating ridge, as we do not see evidence for strongly peaked
sources within the ridge. This is perhaps a matter of interpretation,
although we have found in the past that automated routines sometimes
tend to extract multiple sources in the presence of a ridge of emission 
(Nutter 2004).

SMM2 was seen in the 1.2-mm continuum data taken with the 30-m telescope
of the Institut de Radio Astronomie Millim\'etrique (IRAM)
at 11-arcsec resolution, as reported by
Teyssier et al. (2004) and Habart et al. (2005), although neither of these 
teams discussed this source in particular. We see a broad match between the
IRAM data and the SCUBA data, that are consistent given the different angular
resolutions and noise levels of the different data sets. 

SMM2 also overlaps
with `peak 2' in the C$^{18}$O (J=2$\rightarrow$1) data of Hily-Blant et al. 
(2005), although there is an offset of $\sim$17 arcsec between the peak
positions of our SMM2 and their peak 2. Our 850-$\mu$m peak position
coincides better with the 1.2-mm peak of Teyssier et al. (2004). This offset
is explained by Hily-Blant et al. (2005) as being caused by depletion.
Their peak 1 appears to have been detected in our data, and
the positions of their peaks 3 \& 4 show some 850-$\mu$m dust emission
in our data, but nothing significantly above the
general extended emission of the nebula.
This may be indicating that these latter `peaks'
may simply be generated by optical depth effects in C$^{18}$O.

Johnstone et al. (2005) also detected SMM2 at
850~$\mu$m, and our measured peak flux densities of both SMM1 \& 2
are in good agreement with theirs, so we believe that our flux density
calibration is good. Nonetheless,
we find slightly different extended flux densities in both cases
(even allowing for their two sources in SMM1).
Johnstone et al. (2005) used a complex method to estimate extended
flux densities, that entailed smoothing the data to 130 arcsec and 
subtracting this smoothed image. Then they added back a constant offset
to zero the areas of no emission, and they used an automated routine to
choose their aperture sizes. 

We repeated their technique, and could
only reproduce their measured extended flux densities by using an aperture
that included what we believe to be flux from the extended cloud
as well as from the cores.
Based on our examination of the data, we believe 
our apertures to be a good fit to the source in each case. 
Johnstone et al.
(2005) do not discuss the nature of SMM1 \& 2 in particular, merely
treating them as part of their statistical study of clumps in the Orion
region in general.

SMM2 has not previously been observed in any shorter wavelength observations.
However, close examination of the ISOCAM data in Figure~4 shows that there
appears to be a dip in the emission at the position of SMM2 (c.f. Abergel 
et al. 2003). We believe this could be 
due to absorption of this background emission by the dense core of SMM2.
This dip was also noted by Hily-Blant et al. (2005) and associated
with their peak~2, although it aligns much better with the submillimetre 
source SMM2.

Figure~5 shows a one-dimensional cut of the mid-infrared greyscale shown
in Figure~4 through the centre of SMM2. The cut was made along an axis 
orthogonal to the long axis of SMM2. There is clearly structure associated
with the cloud as a whole. There is possibly a gradient from northwest to 
southeast that can be seen at the extremes of the cut.
However, there is also a very clear absorption `trough' exactly
at the position of SMM2. The horizontal dashed line is an estimate of
the cloud emission away from SMM2, where it appears roughly constant.
We return to this in section 3.2 below, where we
use the depth of the trough to obtain an independent estimate of the column
density of SMM2.

\section{Masses and Densities}

We can use the parameters we have measured for SMM1 \& 2, together with
those measured in previous work, to derive the physical conditions within
the two sources.
Adopting the canonical distance of 400~pc for the Horsehead Nebula (see, for
example, discussion in Pound et al. 2003), we can calculate the actual
sizes of the sources SMM1 \& 2. These are listed in Table 2.
We now treat each source in turn.

\begin{figure}
\setlength{\unitlength}{1mm}
\noindent
\begin{picture}(80,60)
\put(0,0){\includegraphics{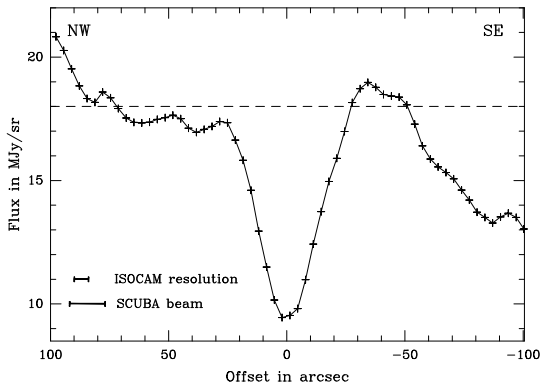}}
\end{picture}
\caption{One-dimensional cut of the 6.75-$\mu$m map shown in Figure~4
through the central position of SMM2, orthogonal to its long axis. The
x-axis is marked in arcsec offset from SMM2, with the north-westerly
direction marked as positive. The y-axis is labelled in units of
MJy/steradian. A clear absorption trough can be seen at the position of 
SMM2. The horizontal dashed line is a fit to the extended emission of
the cloud away from SMM2.}
\end{figure} 

\subsection{B33-SMM1}

Figure 4 shows how the ISOCAM 6.75-$\mu$m emission (Abergel et al. 2003) 
traces the edge of the molecular cloud core SMM1 nearest to the HII region.
As mentioned above, the length of SMM1 is roughly comparable to
the ISOCAM data, although we now see that the width of the submillimetre
source (0.13 pc) is much greater than the width of the ISOCAM filament 
(0.01 pc). We interpreted this above as a temperature gradient across the 
cloud, with the warmer infrared-emitting dust seen along the edge
(which may be due to small grains or PAHs),
while the cooler dust extends deeper into the cloud.
Our submillimetre dust emission appears to match the CO emission
of Abergel et al. (2003) more closely than the mid-infrared ISOCAM emission.
Abergel et al. (2003) derive a kinetic temperature of 30--40~K for the PDR.

\begin{table*}
\caption{Masses and densities derived for B33-SMM1 \& 2. The distance
adopted is 400~pc. The justifications for the adopted distance and dust
temperatures $T$ are
given in the text. Given the assumptions in the mass calculations, the
uncertainties on masses and densities could be a factor
of order a few (see text for discussion). 
Therefore the masses and densities are only quoted to one
significant figure.
The mean column densities $\overline{N}$(H$_2$)
and volume densities $\overline{n}$(H$_2$)
throughout the whole source are quoted, calculated from
the integrated flux densities. The peak column densities N(H$_2$)$_{\rm peak}$ 
and volume densities n(H$_2$)$_{\rm peak}$
are also quoted, as calculated from the peak flux densities.}
\begin{tabular}{lcccccccc}
\hline
Source & Distance & Size & $T$ & Mass &  $\overline{N}$(H$_2$) &
$\overline{n}$(H$_2$) & N(H$_2$)$_{\rm peak}$ & n(H$_2$)$_{\rm peak}$ \\
 & (pc) & (pc) & (K) & M$_\odot$ & (cm$^{-2}$) &
(cm$^{-3}$) & (cm$^{-2}$) & (cm$^{-3}$) \\
\hline
B33-SMM1 & 400 & 0.31 $\times$ 0.13 & 22 & 2   & 4 $\times$ 10$^{21}$ &
1 $\times$ 10$^4$ & 4 $\times$ 10$^{22}$ & 6 $\times$ 10$^5$ \\
B33-SMM2 & 400 & 0.15 $\times$ 0.07 & 15 & 4   & 3 $\times$ 10$^{22}$ &  
1 $\times$ 10$^5$ & 9 $\times$ 10$^{22}$ & 2 $\times$ 10$^6$  \\
\hline
\end{tabular}
\end{table*}

A similar temperature gradient effect is seen by Habart et al. (2005),
who find (from model calculations)
a kinetic temperature of 22~K at a depth of 0.02~pc into the cloud
with gas at a volume density of 2 $\times$ 10$^4$ cm$^{-3}$, and
13.5~K at a volume density of 2 $\times$ 10$^5$ cm$^{-3}$.
We can calculate the mass of gas and dust responsible for emission in
our two submillimetre sources from the 850-$\mu$m flux densities, using
various assumptions, provided we know the temperature of the emitting
dust. This is because
submillimetre continuum emission is usually optically thin, and hence
it is a direct tracer of the mass content of molecular cloud cores
(c.f. Kirk et al. 2005).

For a spherical isothermal dust source
at distance $d$, the total (dust~$+$~gas) mass,
$M(r<R)$, contained within a radius $R$ from the centre,
is related to the submillimetre flux density
$S_{850\mu m}(\theta)$ integrated over a circle of
projected angular radius $\theta = R/d$ by the equation:

\begin{equation}
M(r<R) =  [S_{850\mu m}(\theta)\, d^2]/[ \kappa_{850}\,
B_{850}(T)],
\end{equation}

\noindent
where $\kappa_{850}$ is the dust opacity per unit mass column density
at $\lambda$ = 850$\mu$m
and $B_{850}(T)$ is the Planck function at
the same wavelength, for a dust temperature $T$.

For the dust opacity, we follow the method adopted by
Andr\'e, Ward-Thompson \& Motte (1996) and subsequently used by 
Kirk et al. (2005),
and use $\kappa_{850}$ = 0.01 cm$^2$g$^{-1}$ (see Andr\'e et al. 1996;
Ward-Thompson, Motte \& Andr\'e 1999; and 
Andr\'e, Ward-Thompson \& Barsony 1993;
for detailed justifications both of this value of $\kappa_{850}$
in particular and this method of obtaining masses in general). The
uncertainties in the masses due to a combination of uncertainties in
$\kappa$ could be as high as a factor of a few (see Andr\'e et al. 1996).

For SMM1 we have a range of temperatures which we could adopt, as discussed
above, given that there is probably a temperature gradient across the source.
As stated above,
Habart et al. (2005) found in their model that a mean density of 2 $\times$
10$^4$ cm$^{-3}$ yields a temperature of 22~K, while a mean density of
2 $\times$ 10$^5$ cm$^{-3}$ gives a temperature of 13.5~K. At these
temperatures, the masses derived from submillimetre continuum data
are very sensitive to temperature, due to the
exponential nature of the Planck function.

Adopting a temperature of 22~K gives us a mass for SMM1 of $\sim$2~M$_\odot$,
while a temperature of 13.5~K gives a mass of $\sim$4~M$_\odot$. We can check
for internal consistency by calculating a mean density for SMM1 based on
each of these mases in turn. To do this we need to assume the line-of-sight
dimension of SMM1. Abergel et al. (2003) treat it as an edge-on plane,
whilst Habart et al. (2005) opt for something that is
in fact prolate and filamentary in nature. Hence, we take the average of the 
two and assume that the line-of-sight thickness of SMM1 is the mean  
of its other two dimensions. Then a mass of 2~M$_\odot$ ($T$=22K)
yields a mean volume
number density of n(H$_2$) $\sim$1 $\times$ 10$^4$ cm$^{-3}$, and a mass 
of 4~M$_\odot$ ($T$=13.5K) yields a mean volume
number density of n(H$_2$) $\sim$2 $\times$ 10$^4$ cm$^{-3}$.

Hence we see that a temperature of 13.5~K and a mass of 4~M$_\odot$,
yields a mean volume
density of 2 $\times$ 10$^4$ cm$^{-3}$, compared to the
2 $\times$ 10$^5$ cm$^{-3}$ used by the model to calculate the temperature
of 13.5~K in the first place. Thus this assumption does not yield a
self-consistent result. We note that our line-of-sight assumption only
alters these numbers by a factor of $\sim$2, rather than the 
order-of-magnitude required.

On the other hand,
we see that a temperature of 22~K and a mass of 2~M$_\odot$,
yields a mean volume
density of 1 $\times$ 10$^4$ cm$^{-3}$, which is about equal to
the mean density used by Habart et al. (2005) to calculate the kinetic 
temperature of 22~K in the first place, and the calculation is
self-consistent.
This is also consistent with the mean volume density
derived by Abergel et al. (2003).
Thus the whole derivation is self-consistent and consistent with
previous work, so these are the values we adopt for SMM1, and we list them 
in Table 2. We note that Johnstone et al. (2005) assume a Bonnor-Ebert
form for this source and find a temperature of 19~K, in reasonable
agreement with our adopted value.
We also note that the uncertainties on these numbers could therefore
be a factor of order a few, so we only quote masses and densities to one 
significant figure.

The mean column density we derive for the molecular cloud associated with
SMM1 is 4 $\times$ 10$^{21}$ cm$^{-2}$, which is somewhat higher than the
values derived by Abergel et al. (2003) from their CO data. However, these
latter authors noted that their derived column density was clearly a lower
limit, and only valid for the cloud edge, since they acknowledged that their 
CO data were almost certainly optically thick in the cloud core. 

The peak column density we derive is a factor of 10 higher than the mean 
column density, and corresponds to the densest part of SMM1. The peak volume
density we derive is 6 $\times$ 10$^5$ cm$^{-3}$, although we note a larger
uncertainty in this number due to the unknown line-of-sight depth of the
dense peak region (once again we have assumed that the line-of-sight
dimension of the source is the mean of its other two dimensions).
Nonetheless, this is comparable to the
densities seen in pre-stellar cores (Kirk et al. 2005), which are dense
cores on their way to forming stars (Ward-Thompson et al. 1994).
Hence SMM1 may be pre-stellar in nature, and we discuss this possibility
in section 5 below.

\subsection{B33-SMM2}

The source that we refer to as B33-SMM2 has been detected before at a
wavelength of 1.2~mm (Teyssier et al. 2004; Habart et al. 2005), and also
at 850~$\mu$m (Johnstone et al. 2005). However, most authors have
not paid it much attention. It can also be seen as a minor peak in
the CO(3-2) data (Habart et al. 2005), although again these authors 
concentrated primarily on the PDR rather than other structure in the maps.
Hily-Blant et al. (2005) labelled a nearby CO source as `peak 2' in their 
data, as mentioned in section 2 above, but it is not clear that these
are physically the same source, due to this offset.

There is no shorter wavelength emission detected from SMM2 than our
450-$\mu$m data in Figure 3. In this regard, it is similar to a pre-stellar
core (Ward-Thompson et al. 1994) or a Class 0 protostar (Andr\'e et al.
1993). However, the lack of any known outflow 
emission from this region appears to rule out the latter scenario.

We can calculate the physical parameters of SMM2 in a similar manner to those
we calculated for SMM1. However, once again a large source of uncertainty is 
the dust temperature T within the source. We note that in projection
it is significantly further from the HII region than SMM1 and is
most probably shielded by the bulk of the
B33 molecular cloud, including SMM1. Therefore SMM2 must be cooler than SMM1,
although no previous temperature has been derived for it, since all
molecular line tracers so far observed either do not detect it, or are
depleted or optically thick in this sight-line.

Hence the 22~K we adopt for SMM1 is clearly too high for SMM2. We noted above
the similarity of SMM2 to pre-stellar cores in its extent and appearance
(Kirk et al. 2005). Pre-stellar cores have typical temperatures around 10~K
(Ward-Thompson et al. 2002). However, we note that none of the cores in this
latter study were in the Orion region. Most were in Taurus or Ophiuchus, which
are more quiescent regions, and thus 10~K may be too low for SMM2.

So we have a range of possible temperatures of $\sim$10--20~K.
Therefore we adopt a temperature of 15~K for SMM2, with an error-bar of 5~K.
Johnstone et al. (2005) used a Bonnor-Ebert fit to SMM2 and found a 
temperature of 16~K, in good agreement with our chosen value.
We note that if the temperature were 20~K the derived mass and densities would
decrease by a factor of 2, and if the temperature were 10~K the mass and 
densities would increase by a factor of 1.75.

We derive a mass of 4~M$_\odot$ for SMM2, based on this temperature. Given
that SMM2 is smaller in extent than SMM1, but with higher mass,
it is also therefore more dense. The mean column density is 
an order of magnitude higher than SMM1, as is the mean volume density. 
Note that this difference in densities between SMM1 \& 2
cannot be accounted for simply by our choice of temperatures for
the cores -- this can account for at most a factor of 2 rather than an
order of magnitude.
The peak volume and column densities are also seen 
to be a factor of $\sim$2--3 higher in SMM2 than SMM1. 

However, our calculated mean column density of SMM2 is in very good 
agreement with the 3.5 $\times$ 10$^{22}$ cm$^{-3}$ derived for `peak 2' 
by Hily-Blant et al. (2005), indicating that their peak 2 may in fact be
associated with the extended emission from SMM2. We note that to
calculate the mean volume density we have again assumed that the 
line-of-sight dimension is the mean of the other two. 
Hily-Blant et al. (2005) assumed a cylindrical geometry and found a
mean volume density roughly a factor of 2 lower. But they do not see
the high peak values in their C$^{18}$O data that we see in the dust
continuum. If peak 2 and SMM2 are the same, then this 
may be due to optical depth effects in the CO data, or
to gas depletion onto grains at the highest densities, as they surmised.

We have an independent method of checking our column density estimate for 
SMM2, since as we showed in Figure~5 above, SMM2 is seen as an absorption 
dip against the bright background emission of L1630 in the mid-infrared 
data (c.f. figure~2 of Abergel et al. 2003). From this we can
measure the depth of the absorption at 6.75 and 15~$\mu$m, and
hence calculate the peak column density of SMM2, following the method
of Bacmann et al. (2000).

In this method one compares the intensity of the emission measured at
some position in the general cloud emission that is
away from the core, $I_{off}$, with the intensity of the
emission measured on the core, $I_{on}$. $I_{off}$ is made up of the sum
of the background emission intensity, 
$I_{back}$, from the illuminating source
behind the dense core -- in this case from L1630 (Pound et al. 2003) --
and the widespread general foreground emission intensity, 
$I_{fore}$, that includes
Zodiacal emission and general Galactic emission on scales larger than the
cloud in question. $I_{on}$ is then simply the sum of the foreground
emission intensity
and the attenuated background emission intensity (Bacmann et al. 2000).

So we have

\begin{equation}
I_{off} = I_{back} + I_{fore},
\end{equation}

\noindent
and

\begin{equation}
I_{on} = I_{back}(e^{-\tau_{\lambda}}) + I_{fore},
\end{equation}

\noindent
where $\tau_{\lambda}$ is the optical depth at wavelength $\lambda$.
Rearranging for $\tau_{\lambda}$, gives:

\begin{equation}
e^{-\tau_{\lambda}} = \frac{I_{on} - I_{fore}}{I_{off} - I_{fore}},
\end{equation}

\noindent
and we estimate $I_{fore}$ on a region of the data away from the L1630
cloud emission. There is some uncertainty in the estimate of
$I_{fore}$, since it may vary slightly from one part of the map to
another, and this may be our chief source of measurement error. We attempted
to estimate this by looking at the maximum variation in intensity across
the cloud itself, as well as on regions away from the cloud, and used
this to calculate our errors.

The optical depth can be converted into a column density via the dust
opacity $\sigma_{\lambda}$, using $N(H_2) = \tau_\lambda / \sigma_\lambda$.
We used $\sigma_{6.75\mu m}$ = 1.2 $\times$ 10$^{-23}$ cm$^2$ and
$\sigma_{15\mu m}$ = 1.6 $\times$ 10$^{-23}$ cm$^2$ (Bacmann et al. 2000).

For the 6.75-$\mu$m data we measured
$I_{off}$ = 18 $\pm$ 0.3 MJy/sr, $I_{on}$ = 9.8 $\pm$ 0.1 MJy/sr and 
$I_{fore}$ = 5.5 $\pm$ 0.2 MJy/sr, 
yielding $\tau_{6.75\mu m}$ = 1.07 $\pm$ 0.11.
For the 15-$\mu$m data we measured
$I_{off}$ = 36 $\pm$ 0.5 MJy/sr, $I_{on}$ = 27.5 $\pm$ 0.1 MJy/sr and 
$I_{fore}$ = 25 $\pm$ 0.5 MJy/sr, 
yielding $\tau_{15\mu m}$ = 1.48 $\pm$ 0.34. 

Therefore the 
6.75-$\mu$m data give N(H$_2$) = 8.9 $\pm$ 0.9 $\times$ 10$^{22}$ cm$^{-2}$,
and the
15-$\mu$m data give N(H$_2$) = 9.3 $\pm$ 2.1 $\times$ 10$^{22}$ cm$^{-2}$.
Both estimates are in excellent agreement with our value of peak column 
density for SMM2 in Table 2 of
N(H$_2$) = 9 $\times$ 10$^{22}$ cm$^{-2}$.
We note that no temperature estimate is required in this calculation.
This gives us added confidence in the parameters for SMM2 that we
have calculated from the submillimetre data.

\section{Discussion}

\begin{figure}
\setlength{\unitlength}{1mm}
\noindent
\begin{picture}(80,95)
\put(0,0){\includegraphics{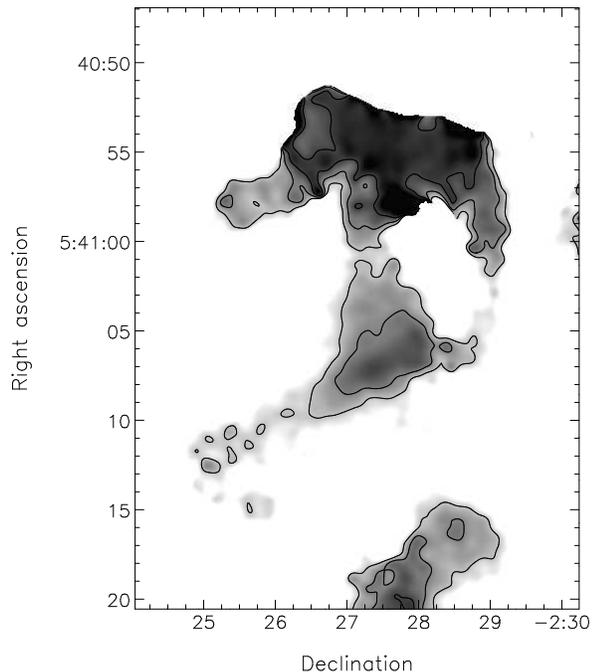}}
\end{picture}
\caption{Image of the ratio of the 850- to 450-$\mu$m images shown in 
Figures 2 \& 3 (after smoothing to a common resolution). Contour levels 
are 0.1, 0.2 and 0.3.}
\end{figure}

We can compare the 850- and 450-$\mu$m data by calculating the ratio of the
emission at these two wavelengths.
Figure 6 shows a ratio map of the data. This was made by first smoothing
the 850-$\mu$m data with 450-$\mu$m beam, and simultaneously smoothing the
450-$\mu$m data with the 850-$\mu$m beam, then subtracting the background 
level from each map and dividing one by the other (after making a cut in
the data to remove low signal-to-noise regions of each map). 
It can be seen that the dust outside of the clumps 
SMM1 \& 2 (that can be seen at high enough signal-to-noise) has a ratio 
of 0.1--0.2 in this image. Higher values are seen towards the cores
of SMM1 \& 2. This is indicating a 
different spectral energy distribution for the two embedded sources
relative to the less dense cloud. In addition there is a difference
between the two cores, such that the majority of SMM1 has a ratio of
0.3, while the majority of SMM2 has a ratio of 0.2.

The submillimetre spectral index $\alpha_{\rm SM}$ is usually defined as:

\begin{equation}
\alpha_{\rm SM} = \frac{log (S_{450}/S_{850})}{log (\nu_{450}/\nu_{850})},
\end{equation}

\noindent
where S$_{450}$ and S$_{850}$ are the flux densities at 450 and 850~$\mu$m
respectively, and $\nu_{450}$ and $\nu_{850}$ are the frequencies at these
wavelengths. A ratio of 0.1 in Figure 6
corresponds to a submillimetre spectral index $\alpha_{\rm SM}$ of
3.6, while 0.2 corresponds to $\alpha_{\rm SM}$ of 2.6,
and a ratio of 0.3 corresponds to $\alpha_{\rm SM}$ of 1.9. 
Values usually observed for dust in molecular clouds range from $\sim$2 
to $\sim$4, as we see here.

Hence we see that the dense sources have a lower value of 
$\alpha_{\rm SM}$ than the rest of the nebula.
This could be caused by a lower temperature in the dense cores
relative to the surrounding lower density cloud,
such as is seen elsewhere in dense cores (Ward-Thompson et al. 2002).
Furthermore, there is a marked difference between SMM1 \& 2, with
the former having a lower value for $\alpha_{\rm SM}$. This cannot be
explained by temperature effects since SMM1 is warmer than SMM2
(see section 3 above). It must be due to different dust
properties in the two cores.

Recent work (Boudet et al. 2005) has shown that in the submillimetre
regime there is an anti-correlation between the spectral index
and the temperature, such that lower spectral indices are seen at higher
temperatures. This work was based on studies carried out of the optical
properties of amorphous silicate grains, but is in agreement with
observations taken by the
Programme National d'Observations Submillimetrique (PRONAOS)
ballon-borne experiment (Lamarre et al. 1994; Lamarre \& Giard 2002). 

The variation of the spectral index seen by PRONAOS (Dupac et al. 2003)
was ascribed to variations in the optical
properties of amorphous silicate grains in different environments in the
inter-stellar medium (Boudet et al. 2005). An alternative explanation
in terms of dust grain coagulation was proposed by Stepnik et al. (2003).
We are possibly seeing one of these effects here.

The Horsehead Nebula is somewhat reminiscent of the Eagle Nebula in
appearance. In both cases columns of dense gas and dust `point' towards
HII regions where massive star formation has taken place. In the case
of the Eagle Nebula, dense clumps are seen at the tips of the columns
(White et al. 1999). For the Horsehead Nebula there is a dense region, SMM1,
at the tip of the column (e.g. Abergel et al. 2003), and we see
that there is another dense clump further `down' the column, SMM2, apparently
further from the ionising radiation of the newly-formed massive stars.
We now discuss the possible formation mechanisms of SMM1 \& 2.

In the Eagle Nebula, White et al. (1999) argued that pre-existing clumps
in the molecular cloud formed barriers to the ionising winds from the
massive stars in the HII region, and that the columns were essentially the 
`wind shadows' of these clumps. In this picture the clumps are eventually
eroded by the massive stars, or else triggered into forming stars, in both
cases meaning they are relatively short-lived. Williams et al. (2001) 
presented an alternative scenario in which columns occur as natural 
instabilities in nature and can be quite long-lived.
For the Horsehead Nebula, Pound et al. (2003) argued for a relatively
short lifetime for the nebula.

Turning to our data, we see that SMM1 is clearly associated with the molecular
cloud behind the PDR that is essentially the `working surface' between the
HII region and the cloud. The shape of the densest part of
SMM1 is somewhat curved, as is the
edge of the obscuration seen at the very `top' of the horse's head
in the optical image. If this curve were seen as an arc that formed
part of a circle, then the centre of that circle would be somewhere in
the vicinity of $\sigma$ Ori, the star believed to be responsible
for forming the PDR (Abergel et al. 2003). This appears to lend support
to the hypothesis that SMM1 has been
affected by its interaction with $\sigma$ Ori.

SMM2 lies further `down' the column of the Horsehead, and apparently
further from $\sigma$ Ori. It also appears to be being shielded by the
remainder of the head and by SMM1. Hence it would appear that SMM2 may
have been a pre-existing clump rather than a density enhancement
caused by some form of `snow-plough effect' (Elmegreen \& Lada 1977).
It is possible that the three-dimensional geometry of the nebula puts
SMM2 slightly foreground to SMM1, and hence it is not shielded. In fact,
in the optical image it is possible to discern a slightly brighter ridge
roughly coincident with the `top' of SMM2, so it may be that it is protruding
somewhat from the near-side of the cloud. However, there is no evidence
for any significant PDR associated with this ridge.

\section{Stability of the condensations}

To determine whether the condensations in the Horsehead are 
prestellar (i.e. destined to spawn stars), we estimate the terms 
in the Virial Theorem, and present them in Table \ref{TAB:VIRIAL}
(c.f. Ward-Thompson 2002).
For a spherical cloud to start contracting, the Virial Theorem 
requires that
\begin{eqnarray}
2\,{\cal U} \,+\, 2\,{\cal T} \,+\, 2\,{\cal R} 
& < & {\cal G} \,+\, {\cal X} \,.
\end{eqnarray}
Here ${\cal U}$ is the thermal energy, ${\cal T}$ is the turbulent 
energy, ${\cal R}$ is the rotational energy, ${\cal G}$ is the 
magnitude of the self-gravitational potential energy, and ${\cal X}$ 
is the contribution from external pressure (${\cal X} = 4\pi R^3 
P_{_{\rm EXT}}$, where $R$ is the cloud radius and $P_{_{\rm EXT}}$ 
is the external pressure). Note that we have ignored the effects of 
any magnetic field that may be present in this region.

Twice the thermal energy is
\begin{eqnarray}
2\,{\cal U} & = & \frac{3\,M\,k_{_{\rm B}}\,T}{\bar{m}} \,,
\end{eqnarray}
where $M$ is the total mass of the clump, $T$ is its temperature,
$\bar{m}$ is the mean molecular weight of the gas and $k_{\rm B}$
is the Boltzmann constant.
We have assumed an isothermal gas which is effectively monatomic 
(since H$_2$ is not rotationally excited). Condensation B33-SMM1 is 
estimated to have mass $M \sim 2\,{\rm M}_{_\odot}$ and temperature 
$T \sim 22\,{\rm K}$, giving $2\,{\cal U} \sim 9 \times 10^{42}\,{\rm erg}$. 
B33-SMM2 has $M \sim 4\,{\rm M}_{_\odot}$ and $T \sim 15\,{\rm K}$, giving 
$2\,{\cal U} \sim 13 \times 10^{42}\,{\rm erg}$. 

\begin{center}
\begin{table}
\caption{Terms in the Virial Theorem for the two clumps in the Horsehead (in 
units of $10^{42}\,{\rm erg}$). If the sum of the first three 
columns is greater than the sum of the final two,
then the cloud will expand. If the sum of the last two columns is greater
then the cloud will collapse.}
\vspace{0.5cm}
\begin{center}
\begin{tabular} {lccccc} \hline
Source & $\;\;\;2\,{\cal U}\;\;$ & $\;\;2\,{\cal T}\;\;$ & 
$\;\;2\,{\cal R}\;\;\;$ & $\;\;\;{\cal G}\;\;$ & 
$\;\;{\cal X}\;\;\;$ \\ \hline
B33-SMM1 & $9$ & $10$ & $3$ & $3$ & $100$ \\
B33-SMM2 & $13$ & $16$ & $1$ & $22$ & $<20$ \\ \hline
\end{tabular}
\end{center}
\label{TAB:VIRIAL}
\end{table}
\end{center}

Twice the turbulent energy is 
\begin{eqnarray}
2\,{\cal T} & = & \frac{3\,M\,\Delta v_{_{\rm FWHM}}^2}{8\,\ln (2)} \,,
\end{eqnarray}
where we have assumed isotropic turbulence, and $\Delta v_{_{\rm FWHM}}$ 
is the full-width at half-maximum of the radial velocity distribution 
(corrected for the thermal contribution). From the optically thin 
C$^{18}$O(2$\rightarrow$1) line observations of Hily-Blant et al. (2005) we 
estimate $\Delta v_{_{\rm FWHM}} \simeq 0.65\,{\rm km}\,{\rm s}^{-1}$, and 
hence $2\,{\cal T} \sim 10 \times 10^{42}\,{\rm erg}$ for B33-SMM1; and 
$\Delta v_{_{\rm FWHM}} \simeq 0.6\,{\rm km}\,{\rm s}^{-1}$, and hence 
$2\,{\cal T} \sim 16 \times 10^{42}\,{\rm erg}$ for B33-SMM2.

Twice the rotational energy is 
\begin{eqnarray}
2\,{\cal R} & = & \frac{4\,M\,R^2\,\Omega^2}{5} \,,
\end{eqnarray}
where $\Omega$ is the angular rotational velocity of the clump.
The values of $\Omega$ for SMM1 \& 2 are estimated from figure~8 of
Hily-Blant et al. (2005).
For B33-SMM1, we adopt $R \sim 0.1\,{\rm pc}$ and $\Omega \simeq 3\,
{\rm km}\,{\rm s}^{-1}\,{\rm pc}^{-1}$, giving $2\,{\cal R} \simeq 
3 \times 10^{42}\,{\rm erg}$. For B33-SMM2, we adopt $R \sim 0.05\,
{\rm pc}$ and $\Omega \simeq 2\,{\rm km}\,{\rm s}^{-1}\,{\rm pc}^{-1}$, 
giving $2\,{\cal R} \simeq 1 \times 10^{42}\,{\rm erg}$.

The magnitude of the self-gravitational potential energy of a
clump is given by
\begin{eqnarray}
{\cal G} & = & \frac{\eta\,G\,M^2}{R} \,,
\end{eqnarray}
where $\eta$ is a coefficient determined by the detailed density
profile of the clump. If we assume that the clump is a critical
Bonnor-Ebert sphere,  which seems to be a good
morphological approximation to many
starless cores, then the coefficient, $\eta$, is given by
\begin{eqnarray}
\eta & = & \left\{ \frac{3}{\xi\,\psi'} \,-\, \frac{{\rm e}^{-\psi}}
{\psi'^2} \right\}_{_{\rm B}} \;=\; 0.732 \,,
\end{eqnarray}
where $\psi(\xi)$ is the standard isothermal function (e.g.
Chandrasekhar 1959), $\psi' \equiv d\psi/d\xi$, and the
right hand side is evaluated at the boundary of the critical
solution (i.e. $\xi_{_{\rm B}} \simeq 6.45$). Using the
values of $M$ and $R$ cited above, we obtain ${\cal G}
\sim 3 \times 10^{42}\,{\rm erg}$ for B33-SMM1, and
${\cal G} \sim 22 \times 10^{42}\,{\rm erg}$ for B33-SMM2.

To calculate the external pressure, we consider the flow of gas 
from a spherical globule which is being eroded by an ionisation 
front. We assume that the gas flows away from the ionisation 
front at constant speed equal to the speed of sound in the 
ionised gas, $a_{_{\rm II}} \sim 10\,{\rm km}\,{\rm s}^{-1}$. 
Consequently the density, $n_{_{\rm II}}(r)$,
in the ionised gas flowing off the 
globule is
\begin{eqnarray}
n_{_{\rm II}}(r) & = & n_{_{\rm II}}(R)\,\left( \frac{r}{R} 
\right)^{-2} \,,
\end{eqnarray}
where $r$ is radius measured from the centre of the globule, 
and $r=R$ is the boundary of the globule. 

Since most of the 
ionising photons incident on the globule will be used up 
maintaining ionisation in the outflowing gas (rather than 
ionising new gas), we can put
\begin{eqnarray} \nonumber
\frac{\dot{\cal N}_{_{\rm LyC}}}{4\,\pi\,D^2} & \simeq & \int_{r=R}
^{r=\infty}\,\alpha_{_\star}\,n_{_{\rm II}}^2(r)\,dr \\ \label{EQN:IONBAL}
 & \simeq & \frac{\alpha_{_\star}\,n_{_{\rm II}}^2(R)\,R}{3} \,.
\end{eqnarray}
Here $\dot{\cal N}_{_{\rm LyC}}$ is the rate at which ionising 
photons are emitted by the exciting star; for $\sigma$ Ori, 
which is an O9.5 star, we adopt $\dot{\cal N}_{_{\rm LyC}} \sim 
3 \times 10^{48}\,{\rm s}^{-1}$ (Schaerer \& de Koter, 1997). 
$D$ is the distance between 
the exciting star and the globule; we adopt the projected 
distances $D\sim 3.5\,{\rm pc}$ for B33-SMM1, and $D \sim 
3.8\,{\rm pc}$ for B33-SMM2. $\alpha_{_\star} \simeq 2 \times 
10^{-13}\,{\rm cm}^3\,{\rm s}^{-1}$ is the recombination 
coefficient for atomic hydrogen into excited states only -- in 
accordance with the On-The-Spot Approximation -- at the canonical 
temperature for the gas in an HII region, $T_{_{\rm II}} \sim 
10^4\,{\rm K}$. 
We note that in reality the outward flow of gas from 
the globule may be accelerated by the inward pressure gradient, 
but the integral in equation~(\ref{EQN:IONBAL}) is dominated by the 
region near the ionisation front, and so this is unlikely to 
be a large correction.

From equation (\ref{EQN:IONBAL}), we obtain
\begin{eqnarray} \label{EQN:IONDEN}
n_{_{\rm II}}(R) & \simeq & \left( \frac{3\,\dot{\cal N}_{_{\rm 
LyC}}}{4\,\pi\,D^2\,R\,\alpha_{_\star}} \right)^{1/2} \,.
\end{eqnarray}
Hence the pressure acting on the boundary of the globule,
$P_B$, is 
given by
\begin{eqnarray}
P_{_{\rm B}} & = & 4\,n_{_{\rm II}}(R)\,k_{_{\rm B}}\,T_{
_{\rm II}} \,,
\end{eqnarray}
where there is a factor 2 to allow for the contribution from 
electrons (assuming $n_{_{\rm e}} \simeq n_{_{\rm p}} \equiv 
n_{_{\rm II}}$) and a factor 2 to allow for the recoil of the 
ionised gas at speed $a_{_{\rm II}}$. Since this pressure only 
acts on one side of the globule, we estimate the contribution 
to the Virial Theorem as
\begin{eqnarray}
{\cal X} & \simeq & 2\,\pi\,R^3\,P_{_{\rm B}} \;\sim\; 
\frac{4\,R^2\,k_{_{\rm B}}\,T_{_{\rm II}}}{D}\,\left( \frac{3\,\pi\,
\dot{\cal N}_{_{\rm LyC}}\,R}{\alpha_{_\star}} \right)^{1/2} \,.
\end{eqnarray}
For B33-SMM1 we obtain ${\cal X} \sim 100 \times 10^{42}\,{\rm erg}$, 
and for B33-SMM2, ${\cal X} \stackrel{<}{\sim} 20 \times 10^{42}\,{\rm erg}$.
We have not taken account of the shadowing effect of the
nebula on SMM2 in this pressure term, 
hence it is quoted as an upper limit in Table~3.

We note that Habart et al. (2005) found an external pressure for SMM1
of $\sim$ 4 $\times$ 10$^6$ Kcm$^{-3}$, which converts to a value
of  ${\cal X} \sim 85 \times 10^{42}\,{\rm erg}$ in our terminology.
This is in remarkably good agreement with our finding of
${\cal X} \sim 100 \times 10^{42}\,{\rm erg}$, using a completely
different calculation. This independent check gives us further
confidence in our calculated values.

From the entries in Table \ref{TAB:VIRIAL}, it appears that neither 
condensation is unequivocally bound if we look at the gravitational
term alone. For SMM1 the left-hand side of equation 6 has a value
of 22 (in units of 10$^{42}$ erg) compared to a gravitational term of only
3. Hence this is roughly an order of magnitude away from gravitational
equilibrium. However, the pressure term has a value of 100, which would
appear to tip the core very strongly in favour of collapse.
We should be mindful that the external pressure only acts on one side of 
the condensation. Therefore it may simply push 
the condensation along by the rocket effect (Kahn 1954; Oort \& Spitzer 
1955) rather than causing it to collapse, although the pressure from
the rest of the cloud would tend to act to prevent this.

For SMM2 the left hand side of equation 6 has a value
of 30 (in units of 10$^{42}$ erg) compared to a gravitational term of 22. 
These two values are remarkably similar, given the uncertainties in all
of the above calculations. For example, an increase in the mass assumed for
SMM2 of $<$20\% is all that is required
make the two values equal. This is well within
the uncertainties of our calculations, as discussed above. Therefore, we
deduce that SMM2 is consistent with being in gravitational virial
equilibrium, without any consideration of the external pressure. 

This would appear to suggest that SMM2 was a pre-existing clump in B33 and is
a candidate pre-stellar core (c.f. Ward-Thompson et al. 2006).
The external pressure term is quoted as an upper limit because we have
not taken account of any shielding of SMM2 by the rest of the cloud.
Nevertheless, if SMM1 is triggered to form stars, and the shock
front and ionisation
front move further into the cloud, then the shielding would be removed 
and the full value quoted in Table~3 would be applicable. In that case we
would say that SMM2 is also destined to collapse under the influence 
of the external pressure.

We re-iterate that inequality 6 is strictly only the
condition for a clump to start contracting. Once
contraction approaches freefall collapse, the non-thermal velocity
dispersion has an increasing contribution from ordered inward
motion, and therefore ${\cal T}$ as defined in equation 8 should be
interpreted as the sum of the turbulent and bulk infall energies,
with the latter contribution becoming dominant. During collapse
$\Delta{\cal T} \simeq \Delta{\cal G}$ and therefore quite quickly
$2{\cal T} > {\cal G}$. Under this circumstance, the only way to
confirm collapse is by measuring asymmetric line profiles.
However, in the present case the fact that for SMM1 $2{\cal U} >
{\cal G}$, and for SMM2 $2{\cal U} \sim {\cal G}$ suggests that
neither clump is collapsing, yet.

We can estimate the speed, $-\dot{R}$,
at which the shock front preceding the ionisation front 
advances into each condensation, using conservation of mass across 
the ionisation front:
\begin{eqnarray} \label{EQN:IFSPEED}
-\,\dot{R} & = & \frac{n_{_{\rm II}}(R)\,a_{_{\rm II}}}
{n(R)} \,;
\end{eqnarray} 
and hence the time $\Delta t$
it will take to ionise the whole of each existing 
condensation, 
\begin{eqnarray} \label{EQN:LIFETIME}
\Delta t & \simeq & \frac{\Delta Z}{(-\,\dot{R})} \,,
\end{eqnarray}
where $\Delta Z$ is the distance to be travelled.
In each case we took the shortest dimension, as we assumed that the
cores are being flattened by the ionisation front, causing them to
lie parallel to it.

Combining equations (\ref{EQN:IONDEN}), (\ref{EQN:IFSPEED}) and 
(\ref{EQN:LIFETIME}), 
we obtain $-\,\dot{R} \sim 0.3\,{\rm km}\,{\rm s}^{-1}$ and 
$\Delta t \sim 0.4\,
{\rm Myr}$ for B33-SMM1; $-\,\dot{R} \sim 0.04\,{\rm km}\,{\rm s}^{-1}$ and 
$\Delta t \sim 1.7\,{\rm Myr}$ for B33-SMM2. However, these 
should be viewed as 
minimum lifetimes for the condensations, since they ignore 
the likelihood that the 
condensations are growing by sweeping up gas on the side 
opposite to the ionisation front.
Nonetheless these values are consistent with the 5-Myr lifetime found
for the entire nebula by Pound et al. (2003).

Hence we find that SMM1 is being strongly
affected by $\sigma$ Ori and hence may be
forced into some form of triggered collapse.
On the other hand SMM2 appears
to have been a pre-existing clump in the molecular cloud --
possibly a pre-stellar core --
which is in approximate gravitational virial equilibrium,
but which may also eventually be forced into collapse by
the external pressure from the HII region.

\section{Conclusions}

We have presented SCUBA images of the Horsehead Nebula, B33,
at 450 and 850~$\mu$m
which show the familiar shape of the dust cloud responsible for the extinction
in this famous nebula. We found strong emission coincident with the molecular
cloud that borders the HII region at the `top' of the horse's head, associated
with the well-studied PDR in this region, which we labelled B33-SMM1.
We calculated the physical parameters of this cloud and found 
that it contains a mass of $\sim$2~M$_\odot$ in a region
0.31 $\times$ 0.13 pc. We also found the density in the heart of 
this region to be as high as $\sim$6 $\times$ 10$^5$ cm$^{-3}$,
which is higher than had previously been seen. We calculated the
virial balance of this clump, and found that the effect of the
ionising radiation from the nearby HII region has the strongest influence 
on its balance, such that it may subsequently
undergo triggered star formation.

In addition we found a source in the `throat' of the horse that looks
just as if it were a lump that the horse
has swallowed. We labelled this source
B33-SMM2. This source has not been previously well studied,
and there appears to be no detection of emission from
this source shortward of our 450-$\mu$m data.
We calculated the mass and density of this source and found
it has a mass of $\sim$4~M$_\odot$ in a region 0.15 $\times$ 0.07 pc,
with a peak density of $\sim$2 $\times$ 10$^6$ cm$^{-3}$. Based on
our virial estimates for this core
we found that SMM2 is in approximate gravitational equilibrium,
consistent with it being a pre-stellar core
that had already formed in B33, but that it may also
eventually be triggered into collapse by the external HII region.

\section*{Acknowledgments}

The James Clerk Maxwell Telescope is operated by the Joint Astronomy 
Centre on behalf of the Particle Physics and Astronomy Research Council 
of the United Kingdom, the Netherlands Organisation for Scientific Research, 
and the National Research Council of Canada.
SCUBA was built at the Royal Observatory, Edinburgh. 
The observations were carried out during the observing run
with reference number M98BN06.
ESO are acknowledged for permission to reproduce Figure~1. 
DN acknowledges PPARC 
post-doctoral support and RA acknowledges PPARC studentship support
that enabled them to work on this project. DWT was on 
sabbatical at the Observatoire de Bordeaux whilst carrying out this work and 
gratefully acknowledges the hospitality accorded to him there.
We also wish to thank Alain Abergel for providing us with the ISOCAM data.

\end{document}